\documentclass[a4paper,11pt]{article}
\usepackage{fullpage}
\usepackage{amsmath}
\usepackage{jheppub}
\usepackage{hyperref}
\usepackage{amssymb}
\usepackage{graphicx}
\usepackage{pgfplots}
\usepackage{tikz}
\usepackage{float}
\usepackage{nicefrac}
\usepackage{physics}
\usepackage{cases}
\usetikzlibrary{patterns}
\usetikzlibrary{calc}

\title{Higher point OTOCs and the bound on chaos}
\author[a]{Pallab Basu,}
\author[a]{Kasi Jaswin}

\affiliation[a]{International Centre for Theoretical Sciences, \\ Tata Institute of Fundamental Research, \\ Bangalore, 560089 INDIA}

\emailAdd{pallab.basu@icts.res.in}
\emailAdd{jaswin@icts.res.in}

\begin{document}
\abstract{ By extending methods of \cite{Maldacena:2015waa}, we investigate the bound on the growth of higher point OTOCs by studying their complex analytical properties. We explore some subtleties in our mathematical investigation, and carefully examine the physical interpretation of our result. We also touch upon the possibility of the saturation of the bound in a physical system. Finally we consider few known examples of higher point OTOCs. For the simplest case of $2n$-point Tremelo correlators the bound on the exponent is proportional to $n$. }
\maketitle	
\section{Commutator, Scrambling and Chaos}
A localized disturbance in a chaotic many body quantum system, with time evolution, spreads over the whole phase space, and information associated with the initial perturbation becomes inaccessible to a simple local measurement. This effect is known as scrambling  \cite{Hayden:2007cs,Sekino:2008he}. In practice scrambling is measured by the growth (decay) of an out of time ordered correlator(OTOC) \cite{Maldacena:2015waa}, 
\begin{align}
F=\expval{V(t)V(0)V(t)V(0)}_\beta.
\label{eq:otoc}
\end{align}
 where in Heisenberg's picture $W(t)=e^{-iHt}W(0)e^{iHt}$ and the expectation is a thermal trace. At an initial time this correlator is finite. However due to scrambling the correlator tends to zero as the time translation operator $U(t)=e^{i H t}$ becomes more convoluted with time. At later time, in a chaotic system, we may replace $U(t)$ by a generic unitary matrix, and assuming maximal scrambling, the correlator may be written as matrix average over all possible unitary matrix,
\begin{align}
F=\int DU \expval{V(t)V(0)V(t)V(0)}
\label{Comm}
\end{align}
and asymptotes to zero in a theory with large number of degrees of freedom.

The OTOC is related to the commutator. In the thermal ensemble the commutator is often zero and one may need to consider the square of it, 
\begin{align}
C&=\expval{[W(t),V(0)]^2}_\beta, \nonumber \\
&=\Tr \left[ e^{-\beta H}[W(t),V(0)]^2 \right]
\label{eq:comm}.
\end{align}
 One may expand the commutator square in \eqref{eq:comm} in two pieces,
\begin{align}
C=\expval{[W(t),V(0)]^2}=-2\expval{W(t)W(t)V(0)V(0)}+2\expval{W(t)V(0)W(t)V(0)}.
\end{align}

The first term in the above expression is a time ordered correlator (TOC) and the second term is an OTOC as in \eqref{eq:otoc}. If we assume something like a large-$N$ factorization, that is if we assume, there is a factorization in the number of degrees of freedom $N_d$, or in another terms the system has a semi-classical description, where fluctuations are small, then we can factorize the TOC as,
\begin{align}
\expval{W(t)W(t)V(0)V(0)} \approx & 2 \expval{W(t)V(0)}^2+\expval{W(t)W(t)}\expval{V(0)V(0)}. 
\end{align}
The first term in the above expression goes to zero at large time due to usual diffusion/relaxation. At a time scale of the order of diffusion time $t_d$, we have, $W(t)V(0) \sim e^{-\frac{t}{t_d}}$.  Hence diffusion with large $N$ factorization gives thermal factorization of TOCs, i.e. all time ordered thermal correlators factorizes to a product of thermal expectations. For our particular example in hand, we have, 
\begin{align}
\expval{W(t)W(t)V(0)V(0)} \approx \expval{W(0)W(0)} \expval{V(0) V(0)}, 
\end{align}
at large time. Hence the large time behavior of the commutator is given by the OTOC \eqref{eq:otoc}. \\ 

At a first look, OTOC may apparently seem to have a similar large $N$ factorization as TOC,
\begin{align}
F\approx 2 \expval{W(t)V(0)}^2+\expval{W(t)W(t)}\expval{V(0)V(0)}+O(1/N_d)
\end{align}
However, the catch is that the sub-leading part of $F(t)$ grows with time. One may argue that for a system with a large number of degrees of freedom, $F$ tends to zero asymptotically. This could be understood as following: in an chaotic system, at an intermediate time much larger than the diffusion time $t \gg t_d$, behavior of $C$ is given by 
\begin{align}
C\propto \epsilon e^{2 \lambda t},
\end{align}
where $\epsilon$ is a small parameter related to the number of degrees freedom and $\lambda$ is the  Lyapunov exponent. Hence, $F$ at an intermediate time would then behave like,
\begin{align}
F \approx f_0-\epsilon f_1 e^{2\lambda t}.
\end{align}
The second term become important at time scale $t_{*}=\frac{1}{\lambda}\log{\epsilon}$, which is known as the scrambling time.

It has been proved in \cite{Maldacena:2015waa} using complex analytic techniques that maximum possible value of Lyapunov exponent has an upper bound proportional to the temperature,
\begin{align}
\lambda_{max} \le \frac{2\pi}{\beta}.
\end{align} 

This maximum value of Lyapunov exponent is also known to saturate in holographic models with gravity (\cite{Shenker:2013pqa,Roberts:2014isa,Shenker:2013yza,Shenker:2014cwa}), certain two dimensional CFTs(\cite{Roberts:2014ifa,Turiaci:2016cvo}) and also in SYK model (\cite{kitaev_vid,Sachdev:1992fk,Maldacena:2016hyu,Polchinski:2016xgd}). 

Recently there are some interest in OTOCs with more than four insertions(\cite{Haehl:2017qfl, Haehl:2017eob,Haehl:2017pak,Haehl:2018izb,Tsuji:2016jbo,Tsuji:2017fxs}., \footnote{Our original motivation to bound higher OTOCs formulated in a discussion with Chethan Krishnan related to his questions about nature of $k$-point OTOCs in a $q$-local SYK model \cite{Krishnan:2016bvg,Krishnan:2017ztz}.} Continuing the same logic as that of the previous paragraph, let us consider the following higher power of the commutator,
\begin{align}
	\label{eq:Cnform}
C=\expval{\prod_{i=1}^{r} [V(T) ,V(0)]^{2n}}.
\end{align}

As $C$ is a product of $2n$-commutators, in a large $N$ theory, it is not difficult to guess how $C$ would behave at large time $T$. It is expected,

\begin{align}\label{commgrowth}
C_n \sim \frac{1}{N^{2r-2}} e^{2 n \lambda  t}.
\end{align}

When expanded, $C_n$ contains and many other OTOCs and the time ordered correlators. In this expansion, $F_n=\expval{(V(T)V(0))^{2n}}$ is the most out of the time ordered OTOC. We intuitively guess $F$ to have highest rate growth (decay) with time compared to other correlators in the expansion of $C_n$. Hence, the large time behavior of $C_n$, determines the large time behavior of $F_n$. \\ 

Here in this work we investigate how analytic properties of an OTOC determines the late time behavior of an OTOC. For that purpose we define a generic correlator which not only captures all possible time orderings in \eqref{eq:Cnform} but also is a function of arbitrary temporal variables. After that, we discussed few known examples in the literature and how our results match with them.

%

\section{On the growth of generic OTOCs }
In this section, we discuss the late time properties of an n-point OTOC. To be well defined a thermal n-point functions needs to be properly regulated. The most generic n-point OTOC with a given scheme of regulation may be expressed as,
\begin{align}
\label{eq:ngencor1}
{\cal F}_{\beta} (t_i, \tau_i) =  \Tr \left( e^{-\beta_1 \hat{H} } V_1(t_1) e^{-\beta_2 \hat{H}} V_2(t_2) \dots e^{-\beta_n \hat{H}} V_n(t_n)  \right)
\end{align}
where, $\hat{H}$ is the Hamiltonian of the system, $\beta_i > 0$ are separations between two consecutive operator insertions along the thermal circle, and therefore satisfy the constraint,  
\begin{align}
\sum_{i=1}^{n} \beta_i = \beta
\end{align} 
where $\beta$ is the inverse temperature of the heat bath i.e the circumference of the thermal circle. ${\cal F}_{\beta} (t_i, \tau_i)$ can visualized as 
\begin{figure}[H]
\centering
\label{fig:Gencr}
\begin{tikzpicture}[scale=1.5, every node/.style={scale=2}]
   \draw (0,0) [smooth,domain=-80:-60] plot ({cos(\x)}, {2*sin(\x)});
   \draw (0,0) [smooth,domain=-58:-20] plot ({cos(\x)}, {2*sin(\x)});
   \draw (0,0) [smooth, domain=-18:-10] plot ({cos(\x)}, {2*sin(\x)});
    \draw (0,0) [smooth, thick, dotted, domain=-10:30] plot ({cos(\x)}, {2*sin(\x)});
   \draw (0,0) [smooth, domain=30:45] plot ({cos(\x)}, {2*sin(\x)});
   \draw (0,0) [smooth, domain=47:90] plot ({cos(\x)}, {2*sin(\x)});
   \draw (0,0) [smooth, domain=90:100] plot ({cos(\x)}, {2*sin(\x)});
   \draw (0,0) [smooth, domain=100:115] plot ({cos(\x)}, {2*sin(\x)});
   \draw (0,0) [smooth, domain=115:120] plot ({cos(\x)}, {2*sin(\x)});
   \draw (-0.5,1.73205) -- (-0.2, 1.73205);
   \draw (-0.544639, 1.67734) -- (-0.2, 1.67734);
   \draw (0,0) [smooth, domain=123:150] plot ({cos(\x)}, {2*sin(\x)});
   \draw (0,0) [smooth, domain=152:180] plot ({cos(\x)}, {2*sin(\x)});
   \draw (0,0) [smooth, domain=182:210] plot ({cos(\x)}, {2*sin(\x)});
   \draw (0,0) [smooth, domain=212:220] plot ({cos(\x)}, {2*sin(\x)});
   \draw (0,0) [smooth, thick, dotted, domain=220:280] plot ({cos(\x)}, {2*sin(\x)});
   \draw (-0.866025,1) -- (-0.1, 1);
   \draw (-0.882948,0.938943) -- (-0.098,0.938943);
	\draw (-0.2, 1.71) [blue, fill=blue] circle (0.043cm);
	\draw (-0.085, 0.97) [blue, fill = blue] circle (0.043cm);
	\draw (-0.999657,-0.0523539) [blue, fill = blue] circle (0.043cm);
	\draw (-0.866025,-1.0) -- (-0.366025,-1.0);
	\draw (-0.848048,-1.05984) -- (-0.348048,-1.05984);
	\draw (-0.338671,-1.035) [blue, fill = blue] circle (0.043cm);
   \draw [-latex] (0,-2.3) to (1,-2.3);
	\draw (0,0) (.25,-2.55) node {$t$};
   \draw (0,0) (0.19, 0.97) node[scale=0.35] {$V_n(t_{n})$};
   \draw (0,0) (0.16, 1.535) node[scale=0.32] {$V_{n-1}(t_{n-1})$};
   \draw (0,0) (-0.7,-0.0523539) node[scale=0.35] {$V_1(t_{1})$};
   \draw (0,0) (-0.038671,-1.035) node[scale=0.35] {$V_2(t_{2})$};
   \draw (0,0) (1.4,0.3) node[scale=1] {$\tau$};
   \draw (0.5,-1.73205) -- (0.7,-1.73205);
   \draw (0.529919,-1.6961) -- (0.70,-1.6961);
   \draw (0.73,-1.71433) [blue, fill=blue] circle (0.043cm);
   \draw (0,0) (1.04,-1.71433) node[scale=0.35] {$V_k(t_{k})$};
   \draw (0.939693,-0.68404) -- (1.3,-0.68404);
   \draw (0.951057,-0.618034) -- (1.3, -0.618034);
   \draw (1.33,-0.651136) [blue, fill=blue] circle (0.043cm);
   \draw (0,0) (1.77,-0.651136) node[scale=0.35] {$V_{k+1} (t_{k+1})$};
   \draw (0.707107,1.41421) -- (0.8, 1.41421);
   \draw (0.681998,1.46271) -- (0.8,1.46271);
   \draw (0.8,1.43868) [blue, fill=blue] circle (0.043cm);
   \draw (0,0) (1.26,1.43868) node[scale=0.35] {$V_{n-2}(t_{n-2})$};
   \coordinate (A) at (1.2,0);
   \coordinate (B) at (1.15,0.65);
   \draw [-latex,smooth] (A) to [bend right=10] (B);
\end{tikzpicture}
\caption{Generic n-point correlator on a thermal circle of radius $\beta$}
\end{figure}
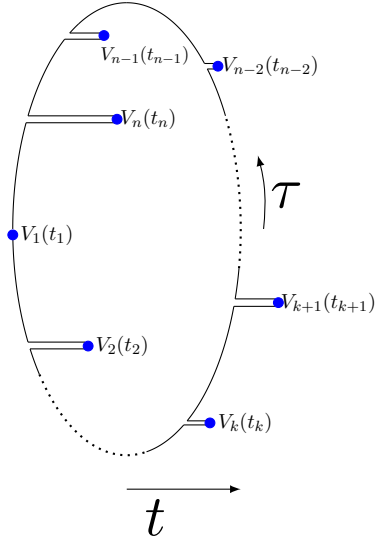
where $V_{i}(t_i)$'s are thermal ordered along $\tau$ direction and time is perpendicular to the thermal circle. One can notice from the figure that the lengths of the insertions are not following ascending or descending order, as the correlator is not time-ordered.
Various properties of the correlator has been discussed in the appendix \ref{ap:gencorrprop}. \\

To proceed we shall assume that each $t_i$, is some function of one temporal parameter $t$, i.e  $t_i = f_i(t)$. In principle $f_i(t)$ can be an arbitrary function of $t$, however in this work we restrict ourselves to increasing linear functions. 
\begin{align}
	t_i \equiv f_i (t) = \omega_i t \qquad \qquad \qquad ; \omega_i \geq 0
\end{align} 
We ask the question, how fast the correlators may grow (decay) with the time parameter $t$. Mathematically it means that we want to put a bound on $\lambda_{l}$, where    
\begin{align}
	\dfrac{\partial{}}{\partial t} |{\cal F}_{\beta}(t)| \leq \lambda_l |{\cal F}_{\beta}(t)| .
\end{align}
 In \cite{Maldacena:2015waa}, the authors considered four-point OTOCs and have shown that 
\begin{align}
	\lambda_{l} \leq \dfrac{2\pi}{\beta}.
\end{align} 
The general idea for deriving a chaos bound for generic n-point correlators can be broken into three steps, which closely follows the derivation given in \cite{Maldacena:2015waa} 
\begin{itemize}
	\item First, find the domain of analyticity of the correlator ${\cal F}_{\beta}$ when time $t_i$  is analytically continued from $t \rightarrow t + i \tau$. For generic n-point correlator it is an asymmetric half-strip.  
	\item Next, find an appropriate normalization factor $\mathcal{N}_\beta$, such that ${g(t)=\cal F}_{\beta}/\mathcal{N_{\beta}} \leq 1$ on the analytic domain. This step needs certain amount of care, because the normalizing factor $\mathcal{N}_{\beta}$ may not be a very large quantity parametrically. 
	
	\item Finally, Schwarz-Pick theorem is used to put a  bound on the growth of the correlator at late times. We find that,
	\begin{align}\label{eq:bound}
	\frac{d}{dt}|g| < (1-|g|) \lambda_l
	\end{align}
	We find that $\lambda_{l}$ is inversely proportional to the width ($\Delta_s$) of the asymmetric strip, 
	\begin{align}
	\label{eq:lyap_width}
	\lambda_l \leq \dfrac{\pi}{\Delta_s}
	\end{align} 
\end{itemize}
 Heuristically we may assume, 
\begin{align}
g\sim g_o-\epsilon e^{\lambda t} g_1,
\end{align}
where $\epsilon$ is a small quantity. Putting the above relation in \eqref{eq:bound}, we can see that the bound becomes a bound on Lyapunov index in a time scale determined by the condition $\epsilon e^{\lambda t} g_1 \gg 1$ or $t \sim \log(\epsilon)-\log(\mathcal{N_{\beta}})$, \eqref{eq:bound} leads to a bound on Lyapunov index. This is why we do not want our normalization factor to be very large. Otherwise it may change the nature of time scale involved. 
 
\subsection{Domain of analyticity}
Domain of analyticity plays an important role in defining the large time behavior of the OTOCs, as it dictates the bound of the correlators. The question is, once analytically continued to complex time variables $t \rightarrow t + i\tau$, what is the domain on which the correlator is well defined. To evaluate that, let's  re-express ${\cal F}_{\beta}$ in \eqref{eq:ngencor1} by introducing dimensionless quantities $\alpha_i = \beta_i/\beta$, and using Heisenberg picture we can express analytically continued ${\cal F}_{\beta}(t+ i \tau, \alpha_i)$ correlator as, 
\begin{align}
	{\cal F}_\beta\qty(t+i \tau,\alpha_i ) & = \Tr \qty[ \rho^{\left(\alpha_1 - \frac{\omega_{n,1}}{\beta} \tau \right)} V_1(t_1) \rho^{\qty(\alpha_2 - \frac{\omega_{1,2}}{\beta} \tau)} V_2(t_2) \rho^{\left( \alpha_3  - \frac{\omega_{2,3}}{\beta} \tau \right)} \ldots \rho^{\left( \alpha_n - \frac{\omega_{n-1, n}}{\beta} \tau \right)} V_n(t_n)]
\end{align}	
where, $\omega_{i,j} = \omega_i - \omega_j$. \\ 
Expressing the trace in the energy basis,
\begin{align}
	{\cal F}_\beta\left(t+i\tau,\alpha_i \right) & = \sum_{m_1, m_2, \cdots, m_n} e^{-\beta E_{m_1}(\alpha_1 - \frac{\omega_{n,1}}{\beta}\tau) } v^{(1)}_{m_1,m_2} e^{-\beta E_{m_2}(\alpha_{i} - \frac{\omega_{1,2}}{\beta}\tau)} v^{(2)}_{m_2,m_3} \cdots  \nonumber \\ 
	& \qquad \qquad \qquad \cdots \;\;  v^{(n-1)}_{m_{n-1},m_{n}}e^{-\beta E_{m_n}({\alpha_n - \frac{\omega_{n-1,n}}{\beta}\tau} ) } v^{(n)}_{m_n,m_{1}}  
\end{align}
where, $v^{(i)}_{a,b}$ are the matrix elements of $V_{i}(t_i)$ operator. Since, energies $E_{a}$ are bounded from below for physically sensible theories, the above summations are convergent as long as the coefficient multiplying $E_{a}$ are positive. When some coefficient, say,  without loss of generality $\alpha_1 - \omega_{n,1} \tau = 0$, then the above expression becomes, 
\begin{align}
	{\cal F}_\beta\qty( t+i \tau, \alpha_i ) & = \sum_{m_2 \cdots, m_n} \sum_{ m_1} \left(v^{(1)}_{m_1,m_2} v^{(n)}_{m_n,m_1} \right) e^{-\beta E_{m_2}\qty(\alpha_{i} - \frac{\omega_{1,2}}{\beta} \tau)} v^{(2)}_{m_2,m_3} \cdots  \nonumber \\ 
	& \qquad \qquad \qquad \cdots \;\;  v^{(n-1)}_{m_{n-1},m_{n}}e^{- \beta E_{m_n}\qty({\alpha_n - \frac{ \omega_{n-1,n}}{\beta} \tau} ) }.   
\end{align}
Now the summation $\sum_{m_1} v^{(n)}_{m_n,m_1} v_{m_1, m_2}^{(1)}$ is not necessarily convergent as it lacks the damping factor $\sim e^{- \# E_{m_1}}$ for higher values of $m_{1}$ \footnote{Here we have assumed discrete spectrum of the Hamiltonian, but similar argument will hold if we consider continuous spectrum as well.}.Therefore the correlator is well defined as long as two neighbouring operators $V_i(t_i)$, $V_{i+1}(t_{i+1})$ don't collide with each other along the thermal circle (they need not be at equal times), and that dictates the domain of analyticity. We find that the domain is given by,    
\begin{align}
\tau_{-} = \text{min. } \left\{  \dfrac{\alpha_j \beta}{\omega_{j,j-1}} \right\} <  \tau  <  \text{min. } \left\{ \dfrac{\alpha_i \beta }{\omega_{i-1,i}} \right\} = \tau_{+}
\end{align}	
Choosing $t_i= 0$ translates to having one $\omega_i = 0$, so the above equation is guaranteed to have solutions and hence ${\cal F}_\beta$ is analytic on a half-strip $\mathcal{D} = (0,\infty)\times i(-\tau_{-}, \tau_{+})$.

It should be mentioned that this is the minimum possible domain of analyticity. Depending on the model, the actual domain could be much bigger. It is expected that in a integrable model the domain would be much wider.  
 
\subsection{Normalization factor}
The derivation of chaos bound through the application of Schwarz-Pick theorem, demands a function bounded by unity on the domain $\mathcal{D}$. This means ${\cal F }_\beta$ needs to be properly after the division by a normalization factor. The normalization factor can be chosen to be greater than the maxima of ${\cal F}_\beta$ on domain $\mathcal{D}$ which is finite and independent of $z \in \mathcal{D}$ coordinates. In an unbounded domain using Phragm\'en-Lindel\"of principle, all one needs show that the function ${\cal F }_\beta$ is bounded by unity on the boundary and is bounded by a constant ${\cal N}_{\beta}$ in the interior. For us the unbounded domain is an asymmetric half-strip of width $\Delta_s$ and $\Re(z) > t_d$, then Phragm\'en-Lindel\"of principle(\cite{pharagmén1908, garrett_2018}) needs that the function in the interior to be less than $\exp(\exp(\frac{\pi}{\Delta_s} \Re (z)))$, i.e if the constant ${\cal N}_{\beta} < \exp(\exp(\frac{\pi}{\Delta_s} t_0))$ then $\qty|{\cal F }_\beta|\le 1$ in the entire domain.

  As we will discuss at the end of this section, there could be some subtleties on how big $C$ could be. This issue is different from the issue of the order of normalization constant and when \eqref{eq:lyap_width} could be interpreted as a bound on the lyapunov index, discussed in the beginning of the chapter. If one is does not need to be careful about this constant $C$ then one can simply use the methods of appendix \ref{ap:traceinq} to put a bound on the correlators to bound a correlator by a product of correlators defined at a higher temperature. In the main text, we will discuss a different method and propose a possible bound on correlators by a product of quantities defined at the same temperature.
 
To find such a bound, we first look at ${\cal F}_{\beta}$ at $(t,\tau_{\pm})$ boundaries. At these boundaries, operators hit each other on the thermal circle and we have various scenario of that happening,
 \begin{itemize}
	\item first only two operators say $V_{k}, V_{k+1}$ hit each other then, ${\cal F}_\beta$ is of the form, 
	\begin{align}
	   {\cal F}_\beta\left(t+i \tau_{\pm},\alpha_i \right) = \Tr \left[  \rho^{b^{\pm}_1} V_1(t_1) \rho^{b^{\pm}_2} V_{2}(t_2) \ldots \left( \rho^{b^{\pm}_{k}} V_k(t_k) V_{k+1}(t_{k+1}) \rho^{b^{\pm}_{k+2}} \right) \ldots \rho^{b^{\pm}_n} V_{t_n}\right].
   \end{align}
   where, 
    \begin{align}
	b^{\pm}_i = \alpha_i \pm \frac{\omega_{i-1,i}}{\beta} \tau_{\pm} 
   \end{align}
   and some $b^{\pm}_{k+1} = 0$. It is not necessary that same two operators hit each other on on upper and lower boundary of the strip. 
	\item Many different pairs of operators hit each other 
	 \begin{align}
		\label{eq:bdycorr}
		{\cal F}_\beta\left(t+i \tau_{\pm},\alpha_i \right)   =  \Tr \left[ \rho^{b^{\pm}_1} V_1(t_1) \rho^{b^{\pm}_2} V_{2}(t_2)  \ldots  
		  \left( \rho^{b^{\pm}_{k}} V_k(t_k) V_{k+1}(t_{k+1}) \rho^{b^{\pm}_{k+2}} \right)  \right. \nonumber \\
		  \left. \ldots \left( \rho^{b^{\pm}_{j}} V_k(t_j) V_{j+1}(t_{j+1}) \right) \ldots \rho^{b^{\pm}_n} V_{t_n} \right]
	\end{align}

 \end{itemize}
and a third scenario where more than two operators hitting each other, which can be avoided by slightly changing the regulation scheme, but not changing the domain of analyticity. Therefore it is sufficient to consider the second scenario \eqref{eq:bdycorr} as the most general case. \\
Suppose $\delta$ is the smallest positive distance between two neighbouring operators, i.e $\text{min. } b^{\pm}_i > 0$, then if we choose some number $k$ such that 
\begin{align}
	\dfrac{1}{2^k} \leq \delta
\end{align}
then we can express the correlator ${\cal F}_{\beta}(t\pm i\tau_{\pm},\alpha_i)$ as
\begin{align}
	\label{eq:Circ_M_form0}
	{\cal F}_{\beta} = \Tr (\prod_{i=1}^{2^k} M_i)
\end{align}
where $M_i$ could have following forms, 
\begin{numcases}{M_i = } 
	\label{eq:Circ_M_form}
	\rho^{\frac{1}{2^{k}}}  \\
	\label{eq:Circ_M_form2}
	  \rho^{\frac{1}{2^k} - \eta} V_i(t_i) \rho^{\eta}  \\
	\label{eq:Circ_M_form3}
  	\rho^{\frac{1}{2^k} - \eta'} V_i(t_i) V_{i+1}(t_{i+1}) \rho^{\eta'} 
\end{numcases}
where $\eta, \eta'$ are some positive numbers less than $\flatfrac{1}{2^{k}}$.
Using the trace inequality identity for product of $2^k$ matrices(proof is given in \ref{ap:trc_bnd_2k_mat}),
\begin{align}\label{eq:trc_bnd_2k_mat}
|{\cal F}_{\beta}| =	
 \qty|\Tr (\prod_{i=1}^{2^{k}} M_i) |  \leq \prod_{i=1}^{2^k} \qty[ \Tr (M_i)^{2^k} ]^{1/2^k}
\end{align}
we can put a bound on ${\cal F}_{\beta}$(diagrammatic demonstration is in \ref{fig:2ksplit})
\begin{align}
	\label{eq:bnd1wnnorm}
\qty|{\cal F}_{\beta} (t\pm i \tau_{\pm}, \alpha_i)| & \leq \prod_{i \; \in \text{ isolated op. }} \qty[\Tr \qty(\rho^{\qty(\frac{1}{2^{k-1}} - 2 \eta_{i} )} V_i \rho^{\qty(2 \eta_{i})} V_i)^{2^{k-1}}]^{1/2^{k}} \nonumber \\ & \quad \times \prod_{j \; \in \text { colliding op.}} \qty[\Tr \qty(\rho^{\qty(\frac{1}{2^{k-1}} - 2\eta_{j}) } V_{j}(t_j) V_{j+1}(t_{j+1}) \rho^{\qty(2 \eta_j)} V_{j+1} (t_{j+1}) V_{j}(t_j))^{2^{k-1}} ]^{1/2^{k}}
\end{align}
trace of operators $M_i$ of the form \eqref{eq:Circ_M_form} will be just one, hence ignored. Index $i$ runs over isolated operator insertions of the form \eqref{eq:Circ_M_form2}, and these traces are independent of time.Index $j$ runs over all colliding operators of the form \eqref{eq:Circ_M_form3}., and we expect them to be finite due to diffusion. For a time $t$ much greater than dissipation time $t_d$, we have following thermal factorization (with a possible error $\varepsilon$),
\begin{align}
	&\qty[\Tr \qty(\rho^{\qty(\frac{1}{2^{k-1}} - 2\eta_{j}) } V_{j}(t_j) V_{j+1}(t_{j+1}) \rho^{\qty(2 \eta_j)} V_{j+1} (t_{j+1}) V_{j}(t_j))^{2^{k-1}} ]  \nonumber \\ 
	&\quad \approx \qty[\Tr \qty(\rho^{\qty(\frac{1}{2^{k-1}} - 2\eta_{j}) } V_{j} \rho^{2 \eta_j} V_{j})^{2^{k-1}} ] \times \qty[\Tr \qty(\rho^{\qty(\frac{1}{2^{k-1}} - 2\eta_{j}) } V_{j+1} \rho^{2 \eta_j} V_{j+1})^{2^{k-1}} ]  + \varepsilon, 
\end{align}
which is a time independent quantity. Hence, $\qty |{\cal F}_{\beta}|$ on the boundary is bounded by a time independent constant, 
\begin{align}
	\label{eq:Normalization}
	 {\cal N}_\beta \equiv \prod_{i \in \text{ all ops. }} \qty[\Tr \qty(\rho^{\frac{1}{2^{k-1}} - 2 \eta_j} V_j \rho^{2 \eta_j} V_{j} )^{2^{k-1}}]^{1/2^k} + \varepsilon 
\end{align}
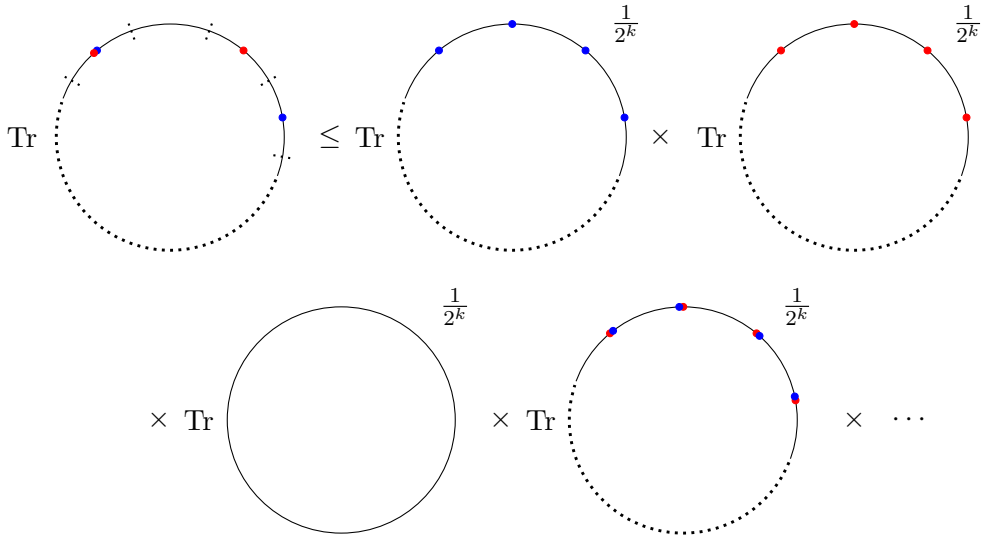
\begin{figure}[H]
	\centering
	\label{fig:2ksplit}
	\begin{tikzpicture}[scale=1.5, every node/.style={scale=1.0}]
	\draw (0,0) circle (1cm);
	\draw (3,0) circle (1cm);
	\draw (6,0) circle (1cm);
	\draw (1.5,-2.5) circle (1cm);
	\draw (4.5,-2.5) circle (1cm);
	\draw (1.4,0) node {$\leq$};
	\draw (4,1) node {$\frac{1}{2^k}$};
	\draw (7,1) node {$\frac{1}{2^k}$};
	\draw (2.5,-1.5) node {$\frac{1}{2^k}$};
	\draw (4.3,0) node {$\times$};
	\draw (2.9,-2.5) node {$\times$};
	\draw (5.5,-1.5) node {$\frac{1}{2^k}$};
	\draw (6.0,-2.5) node {$\times$};
	\draw (6.5,-2.5) node {$\cdots$};
	\draw (-1.3,0) node {$\Tr$};
	\draw (1.75,0) node {$\Tr$};
	\draw (4.75,0) node {$\Tr$};
	\draw (3.25,-2.5) node {$\Tr$};
	\draw (0.25,-2.5) node {$\Tr$};
	\draw (-0.1,-2.5) node {$\times$};
	\draw (0,0) [smooth,line width=5pt,white,domain=-200:-20] plot ({cos(\x)}, {sin(\x)});
	\draw (0,0) [smooth, line width=1.1pt, dotted,domain=-22:-198] plot ({cos(\x)}, {sin(\x)});
	\draw[blue, fill=blue] (0.984808,0.173648) circle (0.03cm);
	\draw[red, fill=red] (0.642788,0.766044) circle (0.03cm);
	\draw[blue, fill=blue] (-0.642788,0.766044) circle (0.03cm);
	\draw[red, fill=red] (-0.669131,0.743145) circle (0.03cm);
	\foreach \x in {-10,30,70,110,150} {
		\draw[thick, dotted] (0,0) -- (\x:1.1);
	}
	\draw (0,0) [white, fill=white ]circle (0.9cm);
	\foreach \x in {10,50,90,130}{
		\draw[blue,fill=blue] ({3+cos(\x)},{sin(\x)}) circle (0.03);
	}
	\draw (0,0) [smooth,line width=5pt,white,domain=-200:-20] plot ({3+cos(\x)}, {sin(\x)});
	\draw (0,0) [smooth,line width=1.1pt,dotted,domain=-22:-198] plot ({3+cos(\x)}, {sin(\x)});
	
	\foreach \x in {10,50,90,130}{
		\draw[red,fill=red] ({6+cos(\x)},{sin(\x)}) circle (0.03);
	}
	\draw (0,0) [smooth,line width=5pt,white,domain=-200:-20] plot ({6+cos(\x)}, {sin(\x)});
	\draw (0,0) [smooth,line width=1.1pt,dotted,domain=-22:-198] plot ({6+cos(\x)}, {sin(\x)});
	\foreach \x in {10,90,50,130}{
		\draw[red,fill=red] ({4.5+cos(\x)},{-2.5+sin(\x)}) circle (0.03);
	}
	\foreach \x in {128,92,48,12}{
		\draw[blue,fill=blue] ({4.5+cos(\x)},{-2.5+sin(\x)}) circle (0.03);
	}
	\draw (0,0) [smooth,line width=5pt,white,domain=-200:-20] plot ({4.5+cos(\x)}, {-2.5+sin(\x)});
	\draw (0,0) [smooth,line width=1.1pt,dotted,domain=-22:-198] plot ({4.5+cos(\x)}, {-2.5+sin(\x)});
	\end{tikzpicture}
	\caption{Diagrammatic demonstration: Top view of thermal circle, blue/red dots indicate operators at different times.}
\end{figure}
For any point $z = t + i \tau$ strictly inside the domain $\mathcal{D}$, if we split the thermal circle into $2^{k}$ segments with operators $M'_{i}(z)$, similar to what is done in \eqref{eq:Circ_M_form0}. If two operators collide in the boundary then we keep them in the same segment. Here, the form of $M'_{i}(z)$'s will be similar to $M_i$ but in the third line \eqref{eq:Circ_M_form3} we will now have some power of $\rho$ inserted between the two colliding operators. For other case \eqref{eq:Circ_M_form2}, only the insertion of $\rho$ will be different from that in the boundary, whereas $M'_i$ ,the operator itself, would be independent of the real part of time ($t$). 

Now in both of cases the ratios between $M'_i(z)$ and $M_{i}$ is only a function of $\tau$ and is of order one. Therefore Applying Phragm\'en-Lindel\"of principle (as discussed in the beginning of the section) if we normalize ${\cal F}_{\beta}(z)$ with ${\cal N}_{\beta}$, it shall remain bounded by unity in the entire domain $\mathcal{D}$.

 In \cite{Maldacena:2015waa} authors have used the contracting properties of $y$ to bound the correlators. This procedure result in an OTOC, where the total sum of the power of $y$ is less than $1$. To express such a correlator as a product of thermal quantities we need to introduce ratios of partition functions defined in two different temperatures. For example in the case of a large-$N$ gauge theory, this results in a bound, which is of order $e^{N^2}$. Hence, technically speaking the particulars of the proof is strictly valid in a time scale of the $o(\log N)+o(\log(1/\epsilon))$. In a large-$N$ gauge theory $1/\epsilon \sim N^2$. This virtually doubles the time scale. To be mentioned is that it is not entirely unreasonable to assume that bounded correlators remain a $o(1)$ number and actual bound is much less what has been proved. If one is fine to use the contracting properties of $y$, then one can use the inequality in the appendix \ref{ap:traceinq} to bound a correlator by a product of correlators defined at a higher temperature. 
 
\subsection{Bound on growth of the correlator}
\begin{figure}[H]
		\label{fig:conftrans}
		\begin{tikzpicture}
			\node[inner sep=0pt] (a) at (-4,0)
				{\includegraphics[scale=0.35]{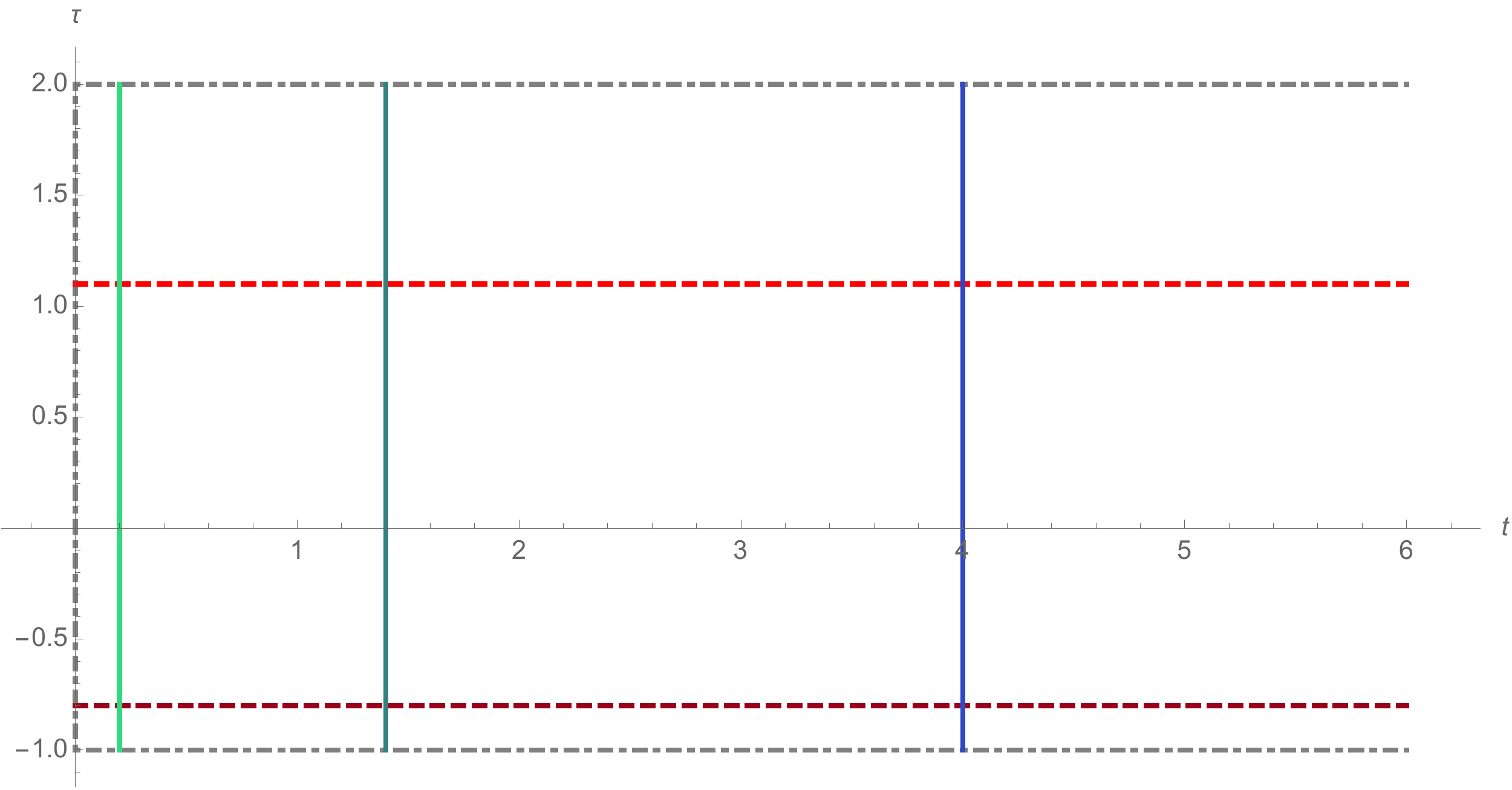}};
				\draw [-latex,smooth,thick] (0.5,0.0) to [bend left=20] (1.5,0.0);
				\node[inner sep=0pt] (b) at (5.5,0)
    			{\includegraphics[scale=0.30]{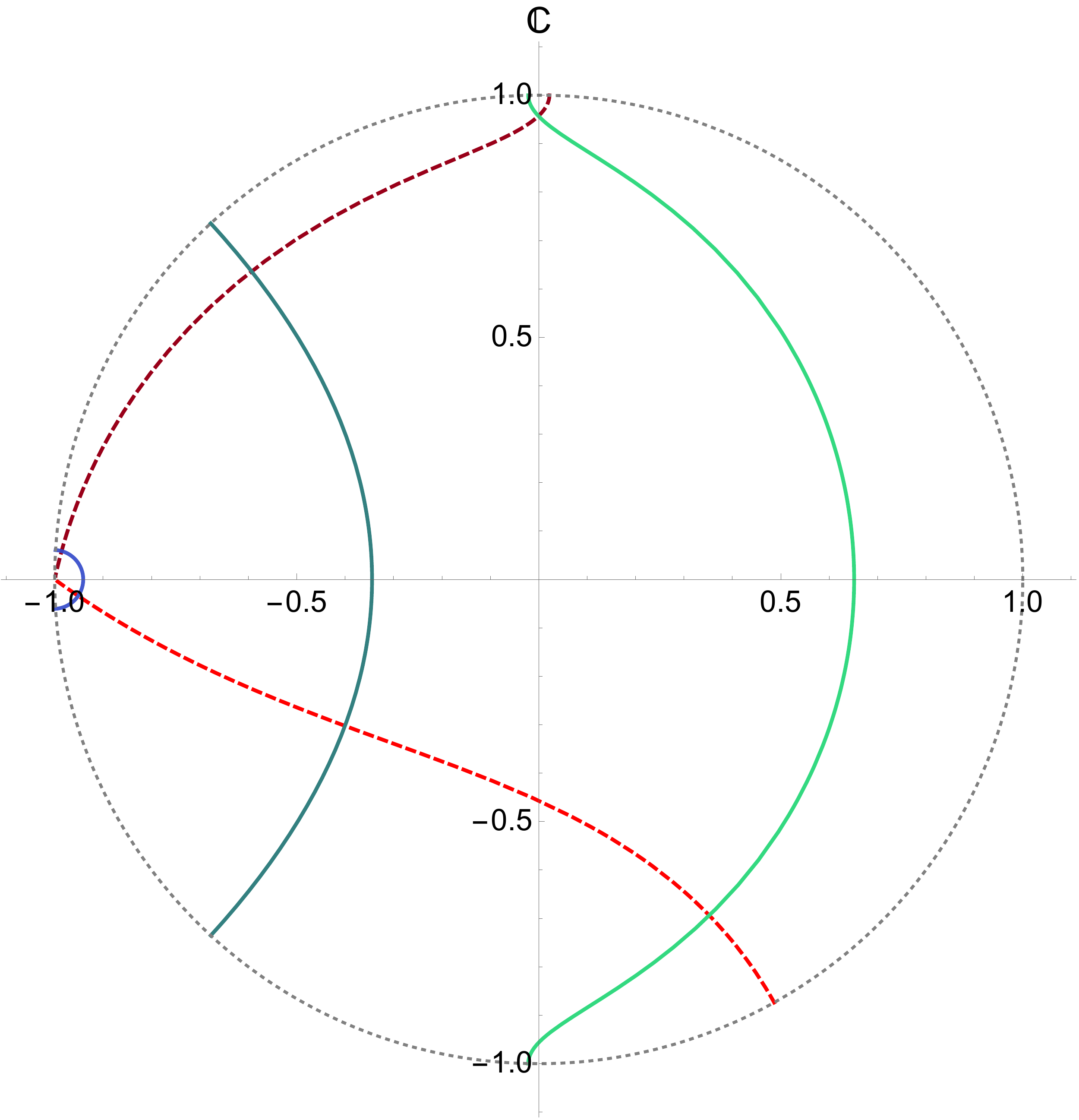}};
			\end{tikzpicture}
	\caption{Conformal mapping from half-strip $\mathcal{D} = (0,\infty) \times i(-1,2) $ to unit disk in complex plane $\mathbb{C}$, each line map to its corresponding colored line-segment on the disk. }
	\end{figure}

Continuing from the previous section $g(t+i\tau) = {\cal F}_{\beta}(t + i \tau)/\mathcal{N}_{\beta} $ is an analytic function and $|g(t + i \tau)| \leq 1$ on the half-strip $\mathcal{D}$. In order to use Schwarz-Pick theorem we conformally map $g$ from $\mathcal{D}$ to a unit disk in complex plane $\mathbb{C}$ using the following transformation  
	\begin{equation}
		z = \dfrac{1- \sinh \left[ \dfrac{\pi}{(\tau_{+} + \tau_{-})}\left(t + i \tau - i\left( \dfrac{\tau_{+}-\tau_{-}}{2} \right)\right) \right]}{1 + \sinh \left[ \dfrac{\pi}{(\tau_{+} + \tau_{-})}\left(t + i \tau - i\left( \dfrac{\tau_{+}-\tau_{-}}{2} \right) \right) \right]}. 
	\end{equation}
	Above mapping is shown in figure \ref{fig:conftrans}, for ($\tau_{+} = 2$, $\tau_{-} = -1$). Fixed-time vertical lines at early times begin as semi-circle on right, with endpoint coordinates $[(0,1),(0,-1)]$ and at late times converge to the point $(-1,0)$ on the unit disk.  

    Let's recall Schwarz-Pick theorem, which states that any conformal mapping $f(z)$ from a unit disk to another unit disk, shall satisfy the following inequality, 
	\begin{equation}
		\dfrac{|df|}{(1-|f(z)|^2)} \leq \dfrac{|dz|}{(1-|z|^2)}
	\end{equation}  
	where $z$ is the coordinates of the unit disk. \\

	Since $g(z) \leq 1$ on $\mathcal{D}$ (as well as on the unit disk), we can think of $g(z)$ as a conformal mapping from a unit disk to another unit disk, and hence applying Schwarz-Pick theorem for a fixed value of $\tau$, we have         
	\begin{align}
		\label{eq:schwarzpick1}
		\dfrac{|\partial_t g(t+i \tau) | }{1-|g(t+i\tau)|^2} \leq \dfrac{\pi}{2\sqrt{2} \Delta_s} \csc \left( \dfrac{\pi (\tau_{+} - \tau)}{\Delta_s}\right) & \text{csch} \left(\dfrac{\pi t}{\Delta_s} \right) \nonumber \\
		& \times \left[ \cos \left( \dfrac{\pi}{\Delta_s} (\tau_{+}-\tau_{-} -2 \tau) \right) +  \cosh \left(\dfrac{2\pi t}{\Delta_s} \right) \right]^{\frac{1}{2}} \nonumber \\
	\end{align}
	where, $\Delta_s$ is the strip width $(\tau_{+} + \tau_{-})$. For $t\gg0$, the above inequality becomes 
	\begin{equation}
		|\partial_t g| \leq \dfrac{\pi}{ 2 \Delta_s} \left[ \csc \left ( \dfrac{\pi}{\Delta_s}(\tau_{+} - \tau) \right)  \right](1- |g|^2) 
	\end{equation}
Expressing $g(t+i\tau) = u(t+i\tau) + i v(t+i \tau)$, where $u,v$ are real-valued functions and using Cauchy-Schwarz inequality one can easily show that 
\begin{align}
	\dfrac{\partial}{\partial t} |g| \leq \left| \dfrac{\partial g}{\partial t} \right|.
\end{align}
Using the above relation, $|g| \leq 1$, and choosing the minima of $\csc(\frac{\pi (\tau_+ - \tau)}{\Delta_s})$ in the domain, the inequality \eqref{eq:schwarzpick1} becomes  
\begin{align}
	\dfrac{d}{d t} |g| \leq \dfrac{\pi}{ \Delta_s}(1- |g|) 
\end{align}
Therefore we have a bound on how ${\cal F}_{\beta}(t)$ decays at late time, 
\begin{align}
\label{eq:mainresult}
|{\cal F}_{\beta}| \lesssim \mathcal{N}_\beta - |{\cal F}_{\beta}(t_0)| e^{\frac{\pi}{\Delta_s} t}
\end{align}
and the Lyapunov exponent can be read-off as $\lambda_L = \dfrac{\pi}{\Delta_s}$ thus, it is just a function of the width of the strip $\mathcal{D}$. 

\section{Examples}
In this section, we find out Lyapunov exponents of few correlators. Before considering higher point correlators let's look at a generalization of four-point OTOC considered in \cite{Tsuji:2016jbo} 
\begin{align}
f_{\gamma} (t) = \dfrac{1}{2} \Tr \qty[ \hat{\rho}^{(1-\gamma)/2} A(t) \hat{\rho}^{\gamma/2} B(0) \rho^{(1-\gamma)/2} A(t) \hat{\rho}^{\gamma/2} B(0)] 	
\end{align}
Using analytic properties of the correlator and assuming late time behaviour of the correlator(briefly described in ap. \ref{ap:gencorrprop}) they conclude that the Lyapunov exponent(i.e $\frac{2 \pi}{\beta}$) is same as the original four-point OTOC. \\
Using our approach we can as well derive the same expected Lyapunov bound, as the domain of analyticity for the above correlator $f_\gamma(t)$ lies between 
\begin{align}
\qty(-\frac{\beta(1- \gamma)}{2}, \frac{\beta \gamma}{2})
\end{align}
and therefore the strip width $\Delta_s = \beta/2$ which is independent of $\gamma$ and giving the same Lyapunov exponent.\\
Next we consider 2n-``Tremolo" correlators mentioned in \cite{Haehl:2017pak} which is has a form $\langle (V(t) W(0))^{n}\rangle$. Choosing a equal-spaced regulation scheme,  
\begin{align}
F^{n}(t) = \Tr \qty( \hat{\rho}^{\frac{1}{2n}} V(t) \hat{\rho}^{\frac{1}{2n}} W(0) )^{n}  
\end{align}
it is easy to see from our method that the Lyapunov exponent $\lambda_l = \frac{n \pi}{\beta}$. Note that, for the above correlator choosing any other type of regulation scheme would give us a greater Lyapunov exponent, therefore equal-space regulation scheme is the best estimate of chaos bound.  
\section{Conclusion}
In this work, we have discussed how complex analytical properties of OTOCs put bound on the temporal behavior of OTOCs. It should be noticed that, in a free or an integrable system, we do not expect to see a chaotic growth of the correlators. It is natural to ask, whether the these bound for a generic operator ever gets saturated in a given theory. As we discussed, such a saturation is known for four point functions in gravity and few other large-$N$ theories.  It is natural to assume that the growth of the higher point  correlators in a black hole back ground also saturates similar bound \cite{Shenker:2014cwa}. In this regard, it would be instructive to have a rigorous calculation of higher point correlators in CFTs and in SYK model or in other computable theories. 

\section{Acknowledgements}
We specially thank Chethan Krishnan for the initiation of the project and initial collaboration. KJ would like to thank Muna Naik and Jitendra Prakash for very useful discussions. PB wants to thank Spenta Wadia and  R. Loganayagam.

\appendix
\section{Properties of Generic n-point correlator}
\label{ap:gencorrprop}
Let's look at few properties of ${\cal F}_{\beta} (t_i, \beta_i)$. In a more conventional way it can be rewritten as, 
\begin{align}
\label{eq:FCtrad}
{\cal F}_{\beta} & = \Tr \left[ \hat{\rho} \left( e^{\tau_1 \hat{H}} V_1(t_1) e^{-\tau_1\hat{H} } \right) \left( e^{ \tau_2 \hat{H}}V_2(t_2)e^{-\tau_2 \hat{H}} \right) \ldots \left( e^{\tau_n \hat{H} } V_{n} (t_n) e^{-\tau_n \hat{H}}\right)  \right ]
\end{align}
where, $\hat{\rho} = e^{-\beta \hat{H}}/\mathcal{Z}$, is the thermal density matrix, \\ 
and $\tau_i \equiv \sum_{j=1}^{i} \beta_j - \beta$. Using Heisenberg picture,  
\begin{align}
A(t) = e^{i\hat{H}t} A(0) e^{-i \hat{H}t}
\end{align}
eqn. \eqref{eq:FCtrad} can be compactly expressed as, 
\begin{align}
{\cal F}_\beta(t_i + i\tau_i) & = \Tr \left[ \hat{\rho} \; V_1\left(t_1 + i\tau_1 \right) V_2 \left( t_2 + i\tau_2 \right) \ldots V_n \left( t_n+ i\tau_n \right) \right] \nonumber \\ 
& = \Tr \left[\hat{\rho} \prod_{i=1}^{n} V_i(t_i + i\tau_i) \right]
\end{align}
where, ${\cal F}_{\beta} (t_i + i \tau_i)$ is understood as ${\cal F}_\beta (t_1 + i \beta_1, t_2 + i\beta_2, \ldots, t_n + i\beta_n)$. \\In this way of writing the correlator, 
\begin{itemize}
\item  the time translation invariance, and trace cyclicity gives rise to following relations,
 \begin{align}
{\cal F}_{\beta}(t_i+i\tau_i+c)&={\cal F}_\beta (t_i+i\tau_i) \\
{\cal F}_\beta (t_i+i\tau_i)&={\cal F}_\beta (t_n+i\tau_n-i\beta,t_1+i\tau_1,\cdots,t_{n-1}+i\tau_{n-1}) 
 \end{align}
 \item  and its complex conjugate
 \begin{align}
 {\cal F}_\beta^{\dagger}(t_1+i\tau_1,\cdots,t_n+i\beta_n)= {\cal F}_\beta (t_n - i \beta_n, t_{n-1} - i \beta_{n-1}, \cdots, t_1 - i \beta_1)
 \end{align}
 implies that the generic correlator is not real, unlike the four-point OTOC considered in \cite{Maldacena:2015waa}.
 \item it is straightforward to see the analytic property  
 \begin{align}
 \left( \dfrac{\partial }{\partial t_{i}} + i\dfrac{\partial}{\partial \tau_i} \right) {\cal F}_{\beta} (t_i + i \tau_i) = 0
 \end{align}
\end{itemize}
Further if we assume, 
\begin{align}
{\cal F}_{\beta}(t)={\cal F}^{0}_{\beta}-\epsilon {\cal F}^{1}_{\beta} e^{\lambda t},
\end{align}
where $t$ is suitable linear combination of $t_i$s. Then the above deferential equation may be used to bound the maximum value of the Lyapunov exponent $\lambda$ \cite{Tsuji:2017fxs}.

\section{Trace Inequalities}
\subsection{Cauchy-Schwarz Inequality for Matrices}
Here we give a short proof of \textit{Cauchy-Schwarz Inequality} for matrices, 
\begin{equation}
\Tr (A^{\dagger} B) \leq [\Tr (A^\dagger A)]^{1/2} [\Tr (B^\dagger B)]^{1/2}
\end{equation}
where $A$, $B$ are some generic finite dimensional matrices. \\ 
\textbf{Proof :} \\ 
Let $[A]_{ij} = a_{ij}$, then $[A^\dagger]_{ij} = a^{*}_{ji}$, similarly for $[B]_{ij}$, we have $b_{ij}$. Then 
\begin{equation}
\Tr (A^\dagger B) = \sum_{i,j} a^{*}_{ki} b_{ki}
\end{equation}
If $\vec{v_i} = (a_{1i}, a_{2i}, \dots)$ and $\vec{w_{i}} = (b_{1i}, b_{2i}, \dots )$, we can write 
\begin{equation}
\Tr (A^\dagger B ) = \sum_{i} \langle v_i | w_i\rangle 
\end{equation}
Using Cauchy-Schwarz(CS) inequality for $\vec{v_i}, \vec{w_i}$ 
\begin{align}
\Tr (A^{\dagger} B) & \leq \sum_{i} ||v_{i} || \;|| w_i ||  \nonumber \\  
& = \sum_i \left( \sum_{k} |a|^2_{ki} \right)^{1/2} \left(\sum_m |b|^2_{mi} \right)^{1/2}   \\  \nonumber
\end{align}
Now thinking of $p_i =  \left( \sum_{k} a^2_{ki} \right)^{1/2}$, $q_{j} = \left( \sum_{k} a^2_{kj} \right)^{1/2}$ as vectors $\vec{p}, \vec{q}$ and using CS inequality 
\begin{align}
\Tr (A^\dagger B) & \leq \langle p | q \rangle \nonumber \\ 
& \leq || p || \; || q || \nonumber \\ 
& = (\sum_{i,k}  |a_{ki}|^2))^{1/2} (\sum_{j,m} |b_{mj}|^2)^{1/2} \nonumber \\ 
& = (\Tr (A^{\dagger} A))^{1/2} (\Tr (B^{\dagger} B))^{1/2} 
\end{align}
Therefore, $ \Tr (A^{\dagger} B ) \leq (\Tr (A^{\dagger} A))^{1/2} (\Tr (B^{\dagger} B))^{1/2} $. 
\subsection{Bound on Trace of Product of Matrices} 
\label{ap:traceinq}
With the help of above inequality, we put on a bound on the trace of product of matrices,
\begin{equation}
	\label{eq:simpl_prd_mat_bd}
	\Tr (\prod_{i=1}^{n} M_i) \leq \prod_{i=1}^{n} \left[\Tr (M^{\dagger}_i M_i ) \right]^{1/2}
\end{equation}
\textbf{Proof : } \\ 
Let, 
\begin{align}
M_{ij} \equiv \prod_{k=i}^{j} V_{k}
\end{align}
where $i \leq j$, and 
\begin{align}
P_i & \equiv V^{\dagger}_i V_i \nonumber \\ 
P_{j,k} & \equiv \left( V_{j} V_{j+1} \dots V_{k} \right)^{\dagger} \left( V_j V_{j+1} \dots V_{k} \right) = M_{jk}^{\dagger} M_{jk}  
\end{align}
notice that, $P$'s are positive semi-definite matrices. \\
Applying CS inequality to $\Tr (V_1 V_2 \dots V_n)$, 
\begin{align}
\Tr (V_1 V_2 \dots V_n) & \leq \left[ \Tr \left( V_1^{\dagger} V_1 \right) \right]^{1/2} \left[ \Tr \left( \left(V_2 V_3 \dots V_n \right)^{\dagger} (V_2 V_3 \dots V_n) \right)\right]^{1/2} \nonumber \\ 
& = \left[ \Tr (P_1) \right]^{1/2} \left[ \Tr \left( P_{2,n} \right)\right]^{1/2} 
\end{align}
to the above inequality, applying CS inequality to $\Tr \left(P_{2,n} \right)$, we have 
\begin{align}
\Tr (V_1 V_2 \dots V_n) & \leq  \left[ \Tr (P_1) \right]^{1/2} \left[ \Tr \left( P_{n} \right)^2 \right]^{1/4} \left[ \Tr \left( P_{2,n-1} \right)^2 \right]^{1/4} \nonumber \\ 
& \leq \left[ \Tr (P_1) \right]^{1/2} \left[ \Tr \left( P_{n} \right) \right]^{1/2} \left[ \Tr \left( P_{2,n-1} \right) \right]^{1/2}
\end{align} 
in the second line we have used the fact that, $\left[\Tr (P^2) \right]^{1/2} \leq \Tr (P)$ for positive semi-definite matrices. Now recursively applying CS inequality to $\Tr (P_{2,k})$'s and using norm-inequality we get the desired bound \eqref{eq:simpl_prd_mat_bd}. \\ 
Using the above inequality, the correlator ${\cal F}_\beta$ in \eqref{eq:bdycorr} is bounded by 
\begin{align}
	\label{eq:bndonstrp}
	{\cal F}_\beta\left(t+i \tau_{\pm},\alpha_i \right)  \leq \prod_{\{ i \in \text{ single op.}\} } &\left[\Tr (\rho^{b^{\pm}_i} V_i \rho^{b^{\pm}_{i+1}} V_{i})\right]^{1/2} \nonumber \\ & \prod_{\{ j \in \text{ double op.} \}}\left[ \Tr \left( \rho^{b^{\pm}_j} V_{j}(t_j) V_{j+1}(t_{j+1}) \rho^{b^{\pm}_{j+2}} V_{j+1}(t_{j+1}) V_{j}(t_{j}) \right) \right]^{1/2} 
\end{align}
here index $i \in \{ \text{ single op.} \}$ runs over all $P_{i}$ with single operators of the form as in \eqref{eq:Circ_M_form2}, and index $  j \in \{ \text{ double op.} \}$ runs over $M_{j}$ of the form given in \eqref{eq:Circ_M_form3} . It is easy to see that the product of single operator traces are independent of time, but not double operator traces. To have a time-independent normalization factor, we need to find an upper bound on the time dependent part of the above inequality. For that, notice that the individual double operator trace is in fact time ordered, and now setting the initial time $t_0$ much greater than the dissipation time $t_d$ but much less the scrambling time $t_{*}$, \eqref{eq:bndonstrp} these operators factorize to give a time-independent bound, 
\begin{align}
	\label{eq:actBndonstrp}
	{\cal F}_\beta\left(t+i \tau_{\pm},\alpha_i \right) & \leq  \prod_{\{ i \in \text{single op.} \} } \left[\Tr (\rho^{b^{\pm}_i} V_i \rho^{b^{\pm}_{i+1}} V_{i})\right]^{1/2} \nonumber \\ &  \qquad  \qquad \quad \times \prod_{\{ j \in \text{ double op.} \}}\left[ \Tr \left( \rho^{b^{\pm}_j} V_{j}\rho^{b^{\pm}_{j+2}} V_{j} \right) \right]^{1/2} \left[ \Tr \left( \rho^{b^{\pm}_j} V_{j+1} \rho^{b^{\pm}_{j+2}} V_{j+1} \right) \right]^{1/2}  
\end{align}

\subsection{Trace Inequality for product of $2^{k}$ Matrix}
\label{ap:trc_bnd_2k_mat}
The above inequality is useful for arbitrary number of matrices, but if the number of matrices is a power of $2$, say $2^{k}$ then there is an interesting inequality 
\begin{align}
	\abs{\Tr (\prod_{i=1}^{2^k} M_i) } \leq \prod_{i=1}^{2^k} \qty[\Tr (M_i)^{2^{k}}]^{1/2^k}
\end{align}
\textbf{Proof : } \\ 
Starting with CS inequality for two matrices $A_{1,2}$
\begin{align}
	\qty|\Tr(A_1 A_2)| & \leq \left( \Tr (A_1^{\dagger} A_1 )  \Tr(A_2 A_2^{\dagger}) \right)^{1/2} = \left( \Tr (A_1^{2}) \Tr(A_2^{2}) \right)^{1/2}
\end{align}
now if $A_i$ is itself a product of two matrices $B_{2i-1} B_{2i}$, then the above inequality becomes  
\begin{align}
	\qty|\Tr (B_1 B_2 B_3 B_4)| & \leq \qty(\Tr (A_1^2 ) \Tr \qty(A_2^2))^{1/2} \nonumber \\ 
	& = \qty[\Tr \qty(B_1^2 B_2^2) \Tr (B_3^2 B_4^2) ]^{1/2} \nonumber \\ 
	& \leq \qty[ \Tr \qty(B_1^{4}) \Tr (B_2^4)  \Tr (B_3^4) \Tr (B^{4}_4)]^{1/4}
\end{align}
hence for using the above argument recursively for a product of $2^{k}$ matrices $M_i$, we can show that 
\begin{align}
	\abs{\Tr (\prod_{i=1}^{2^k} M_i) } \leq \prod_{i=1}^{2^k} \qty[\Tr (M_i)^{2^{k}}]^{1/2^k}.
\end{align}

\bibliography{higherotoc}{}
\bibliographystyle{unsrt}
\end{document}